\documentclass[10pt,conference]{IEEEtran}
\pdfoutput=1
\IEEEoverridecommandlockouts
\usepackage{cite}
\usepackage{amsmath,amssymb,amsfonts}
\usepackage{algorithmic}
\usepackage{graphicx}
\usepackage[caption=false]{subfig}
\usepackage{textcomp}
\usepackage{xcolor}
\usepackage{color, soul}
\usepackage{enumitem}
\usepackage{import}
\usepackage[ruled,vlined]{algorithm2e}
\usepackage{setspace}
\def\BibTeX{{\rm B\kern-.05em{\sc i\kern-.025em b}\kern-.08em
    T\kern-.1667em\lower.7ex\hbox{E}\kern-.125emX}}
\begin{document}

\title{DERauth: A Battery-based Authentication Scheme for Distributed Energy Resources
}

\author{\IEEEauthorblockN{\textbf{Ioannis Zografopoulos, Charalambos Konstantinou}}
\IEEEauthorblockA{Department of Electrical and Computer Engineering, FAMU-FSU College of Engineering \\
Center for Advanced Power Systems, Florida State University\\
Email:\{izografopoulos, ckonstantinou\}@fsu.edu}}

\maketitle

\begin{abstract}
Over the past decades, power systems have experienced drastic transformations in order to address the growth in energy demand, reduce carbon emissions, and enhance power quality and energy efficiency. This shift to the smart grid concept involves, among others,  the utilization of distributed energy resources (DERs) such as rooftop solar panels and storage systems, contributing towards grid decentralization while improving control over power generation. In order to seamlessly integrate DERs into power systems, embedded devices are used to support the communication and control functions of DERs. As a result, vulnerabilities of such components can be ported to the industrial environment. Insecure control networks and protocols further exacerbate the problem. Towards reducing the attack surface, we present an authentication scheme for DERs, \textit{DERauth}, which leverages the inherent entropy of the DER battery energy storage system (BESS) as a root-of-trust. The DER authentication is achieved using a challenge-reply mechanism that relies on the corresponding DER's BESS state-of-charge (SoC) and voltage measurements. A dynamically updating process ensures that the BESS state is up-to-date. We evaluate our proof-of-concept in a prototype development that uses lithium-ion (li-ion) batteries for the BESS. The robustness of our design is assessed against modeling attacks performed by neural networks.
\end{abstract}


\graphicspath{ {ISVLSIFigures/} }

\section{Introduction} \label{s:introduction}
According to North American Electric Reliability Corporation (NERC), a distributed energy resource (DER) is any resource on the distribution power grid that generates electricity and is not otherwise included in the bulk electric system \cite{NERCDER}. 
DERs include microgrids, energy storage, behind-the-meter generation, etc. The increasing amounts of DERs -- their generation capacity is expected to be 40GWs by 2030 \cite{DER_outlook} -- contribute towards the transformation of the energy infrastructure offering flexible control over power generation while minimizing operating costs. Specifically, DERs with energy storage (e.g., fuel cells, batteries, or flywheels) extend grid reliability while efficiently addressing the balance between real-time energy supply and demand. Battery energy storage systems (BESS) contribute significantly to this balancing process. Among different options, lithium-ion (li-ion) batteries have become the dominant form for BESS installations due to their high energy density, decreasing costs, and performance characteristics \cite{hall2008energy}.

In order to effectively monitor and control the operation of DERs, DER plants involve both local communications between the plant management system and DER units as well as between the plant and operators or aggregators who ``manage the DER plant as a virtual source of energy and ancillary services'' \cite{DERComm}. 
The interconnection and interoperability of DERs are  enabled by protocols such as Modbus, DNP3, and IEEE 2030.5. However, such interfaces do not typically include any normative and overarching cybersecurity requirements. For example, Modbus and DNP3 have several identified vulnerabilities \cite{TAXnon}. 
The implementation of these protocols is often enabled by legacy embedded systems developed without security in mind. This is evident by the 
incidents against the Ukrainian grid targeting embedded devices 
supporting industrial communications \cite{Ukr1, konstantinou2017security}.

Towards addressing the aforementioned security issues, several software and hardware approaches have been developed \cite{mclaughlin2016cybersecurity}. Most software-based monitoring techniques add instrumented code into the original application \cite{burow2017control}. Their computing overhead, however, is too large to be deployed in practice, especially in real-time applications \cite{basu2019preempt, edseee.646149920130301}. In addition, software-based protocols can be exploited due to implementation vulnerabilities, enabling network attacks or even allow adversaries to disable the protocol communication media \cite{faruk2008testing}. 

\begin{figure}[t!]
\centerline{\includegraphics[width=\linewidth]{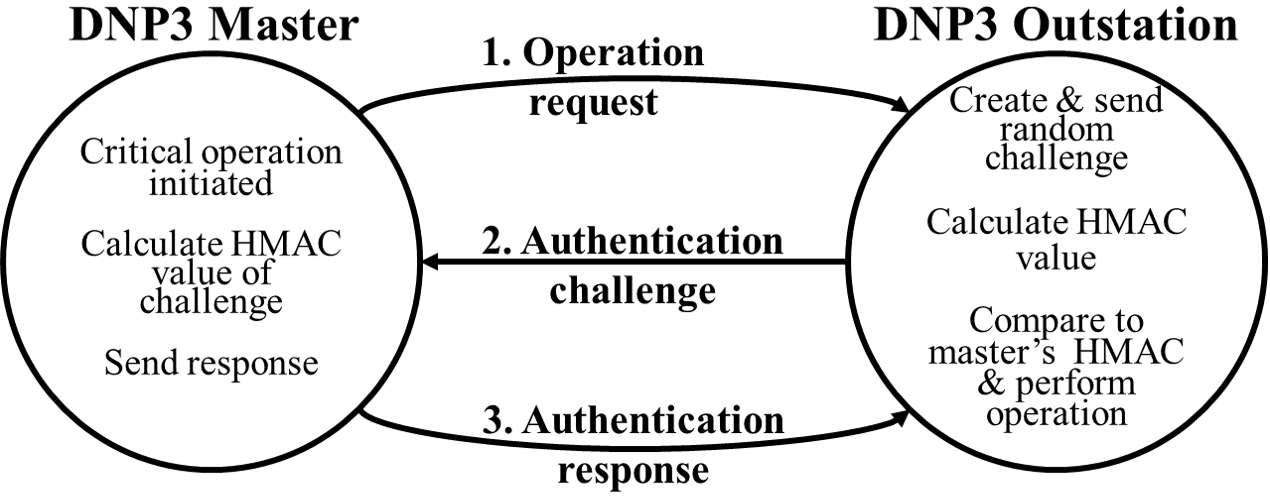}}
\caption{Secure CRSeq authentication of DNP3-SA, a communications protocol used between components in process automation and industrial systems.}
\label{fig:dnp3sa}
\end{figure}
\begin{figure*}[t]
\centerline{\includegraphics[width=0.8\linewidth]{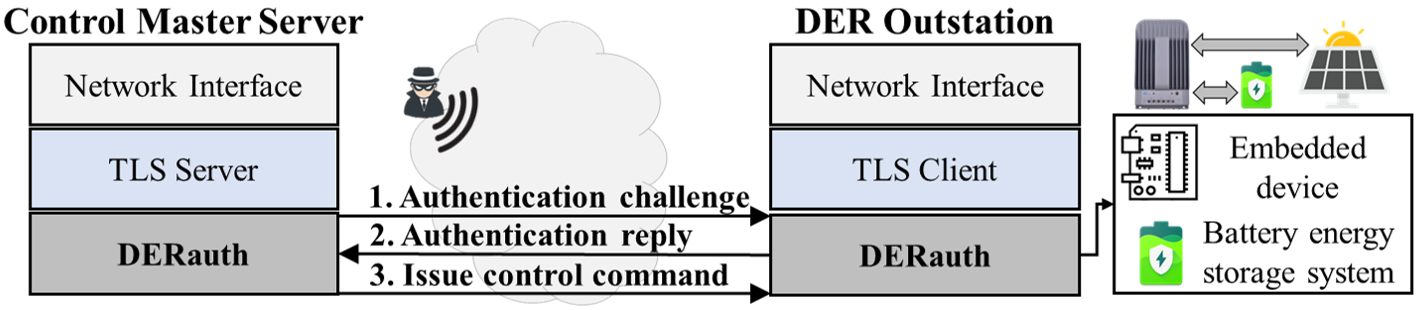}}
\caption{Overview of the communication configuration with DERauth between DER system operators and DER units.}
\label{fig:protocol_stack}
\end{figure*}

Methods that utilize hardware as a root-of-trust can provide a firm foundation from which to build security and trust \cite{hardwareT}. Hardware-assisted technologies include trusted platform modules (e.g., Intel TXT), trusted execution environments  (e.g., ARM TrustZone), virtual isolation (e.g., Intel VT), cryptographic acceleration (e.g., Intel AES-NI), random number generation (RNG) (e.g., SRAM physical unclonable functions -- PUFs), etc. For instance, PUFs leverage the physical properties of hardware devices to extract randomness ensuring unclonability, prediction infeasibility, and tamper-evident properties.

In this paper, we present \emph{DERauth}, a lightweight hardware-based authentication scheme which can be implemented by deployed embedded devices such as programmable logic controllers and gateways (located at DERs and aggregators)  and serve as an add-on feature in existing protocols.  DERauth provides a secure method for plant and fleet operators, utilities, retail energy providers, and aggregators to authenticate individual DER systems at various facilities by utilizing the unique hardware characteristics of DERs to serve as a root-of-trust. Authentication support features of existing communication protocols used in power grid substations typically rely on challenge-reply sequences (CRSeqs) \cite{DNP3-SA}. Such schemes assist in verifying that all received commands within the power grid are genuinely sent by authorized remote embedded devices. An example of the CRSeq authentication as part of the secure authentication extension of the distributed network protocol (DNP3-SA, IEEE 1815-2012 std.) is shown in Fig. \ref{fig:dnp3sa}.

The functionality of DERauth is based on the concept of CRSeq authentication making it applicable to a wide range of industrial protocols. Instead of relying on software generated values (e.g., message authentication codes -- MACs), DERauth leverages the physical characteristics of DER assets. Fig. \ref{fig:protocol_stack} presents an overview of the approach within the typical communication configuration between operators and DER plants. 
DERauth utilizes the properties of li-ion BESS to extract randomness using the BESS real-time state consisting of state-of-charge (SoC) and voltage measurements. 
Specifically, we \textit{(i)} introduce a BESS-based self-authentication scheme for DERs, \textit{(ii)} utilize a dynamically updating process to improve reliability and account for BESS aging and cycle-to-cycle variations, and \textit{(iii)} address  modeling attacks by incorporating BESS data into DERauth's replies.

The rest of the paper is organized as follows. Section II presents the background and related work. Section III describes our methodology while Section IV presents the 
experimental results. Section V concludes the paper. 

\section{Background And Related Work} \label{s:Background And Related Work}

Most industrial protocols 
were initially developed without any security features, i.e., they do not support mechanisms to ensure data integrity and confidentiality. They rely on an \emph{a priori trusted} relationship between master and slave communication devices. To overcome these pitfalls, security measures involve software assisted wrapper functions. However, such methods have been shown to be susceptible to a variety of attacks since it is easy to decode the algorithm and extract secret keys, especially if stored in the non-volatile memory of the device \cite{DNP3-SA, shamsoshoara2019survey}.

The proliferation of hardware-assisted security solutions assumes that hardware can inherently be trusted as the lower lever of abstraction, and thus contribute in reducing the attack surface of embedded devices within industrial environments \cite{HardSec2019, wang2016malicious, konstantinou2018phylax}. Among them, PUFs are lightweight security structures relying on a challenge-response mechanism where for each challenge provided as input the PUF reacts in a unique and unpredictable (but \emph{repeatable}) way. PUFs can be classified based on their fabrication method as silicon (e.g., SRAM and Arbiter PUFs) and non-silicon PUFs (e.g., MEMS and piezo-sensor PUFs) \cite{allPUFs}. 
Silicon PUFs, however, need to be designed during the integrated circuit (IC) fabrication process \cite{sram,arbiter}. On the other hand, non-silicon PUFs require instrumentation and quantization schemes \cite{willers2016mems, piezo}. Existing solutions cannot be integrated in industrial environments with already deployed embedded devices and established communication protocols since redesigning or suspending system operation can be intolerable \cite{falas2019hardware}.  DERauth relies on existing infrastructure to provide hardware-based authentication support for industrial protocols.

Since many protocols that have been used for remote operation of industrial assets did not originally support any security properties, and as industrial environments are less likely to be supported by a dedicated infrastructure, such protocols started to incorporate add-on security features. An example is the hash-based MAC (HMAC) authentication of DNP3-SA (Fig. \ref{fig:dnp3sa}). A DNP3 master device sends an operation request to the outstation device which, upon receiving a critical request,  sends an authentication challenge message to the master. The master device calculates the HMAC for the challenge and sends it back to the outstation that computes the HMAC value for the challenge message and compares it with the received one. If the values match, the DNP3 outstation executes the operation request. Despite the security mechanisms of DNP3-SA, the protocol has still significant 
drawbacks such as: \textit{(i)} the utilization of HMAC-SHA-1 as its MAC algorithm that has been proven to be cryptographically weak, and \textit{(ii)} the dependence on pseudo-RNGs (PRNGs) for its session keys in absence of a high-entropy source \cite{SHA, DNP3-SA}.

\section{{DERauth} Design}  \label{s:methodology}

DERauth aims to address the security issues of industrial protocols by leveraging the physical properties of DERs while minimizing redesign efforts. We follow a similar approach to DNP3-SA, in which the execution of an operation request requires authentication of the outstation DER embedded device using a CRSeq. To achieve that, we leverage the inherent physical randomness from the BESS using SoC and voltage measurements as the sensed physical quantities to construct \emph{non-repeatable} replies. Thus, for instance, if DERauth is used with DNP3-SA it would alleviate the issue of relying on PRNGs to build session keys for HMAC. 
In order to address modeling attacks \cite{Modelling, delvaux2019machine},  we compare the BESS measurements of distinct batteries with their previous cycle measurements (i.e., \emph{self-authentication}) rather than performing cell-to-cell comparisons. We also  incorporate the BESS real-time measurements in our replies to increase entropy. The latter ensures that same challenges always result in  \emph{different} replies, hence mitigating eavesdropping and packet replay attacks. Furthermore, instead of using HMACs to validate the integrity of exchanged data, DERauth includes a transformation function that cannot only support similar HMAC functionality but also assists in mitigating modeling attacks.

\subsection{Threat Model}
The objective of DERauth is to ensure that a \emph{control master server}, either at the DER plant management system or the utility, is able to authenticate a \emph{DER outstation unit} in the presence of an attacker eavesdropping the communication channel. The embedded outstation device is placed at a secure location and can acquire real-time BESS measurements. In addition, the master device as part of a secure energy control facility is operated only by authorized users who enforce security mechanisms to mitigate risks related to operational disruptions. We consider the Dolev-Yao threat model where any communication channel between two parties is considered insecure after the initial handshake (\emph{enrollment})  \cite{nicanfar2013multilayer}. The Dolev-Yao adversary is capable to perform man-in-the-middle 
attacks and inject, eavesdrop, modify, and block messages on the communication network in order to get authenticated. Furthermore, DERauth can handle adversaries knowing protocol-specific information. We also consider modeling attacks in which the adversary acquires CRSeqs (e.g., by eavesdropping) aiming to reconstruct the challenge-reply mechanism of DERauth and initiate a communication request as a trusted DER unit.

\subsection{Li-ion Battery Cells}

\begin{figure}[t]
\centerline{\includegraphics[width=\linewidth]{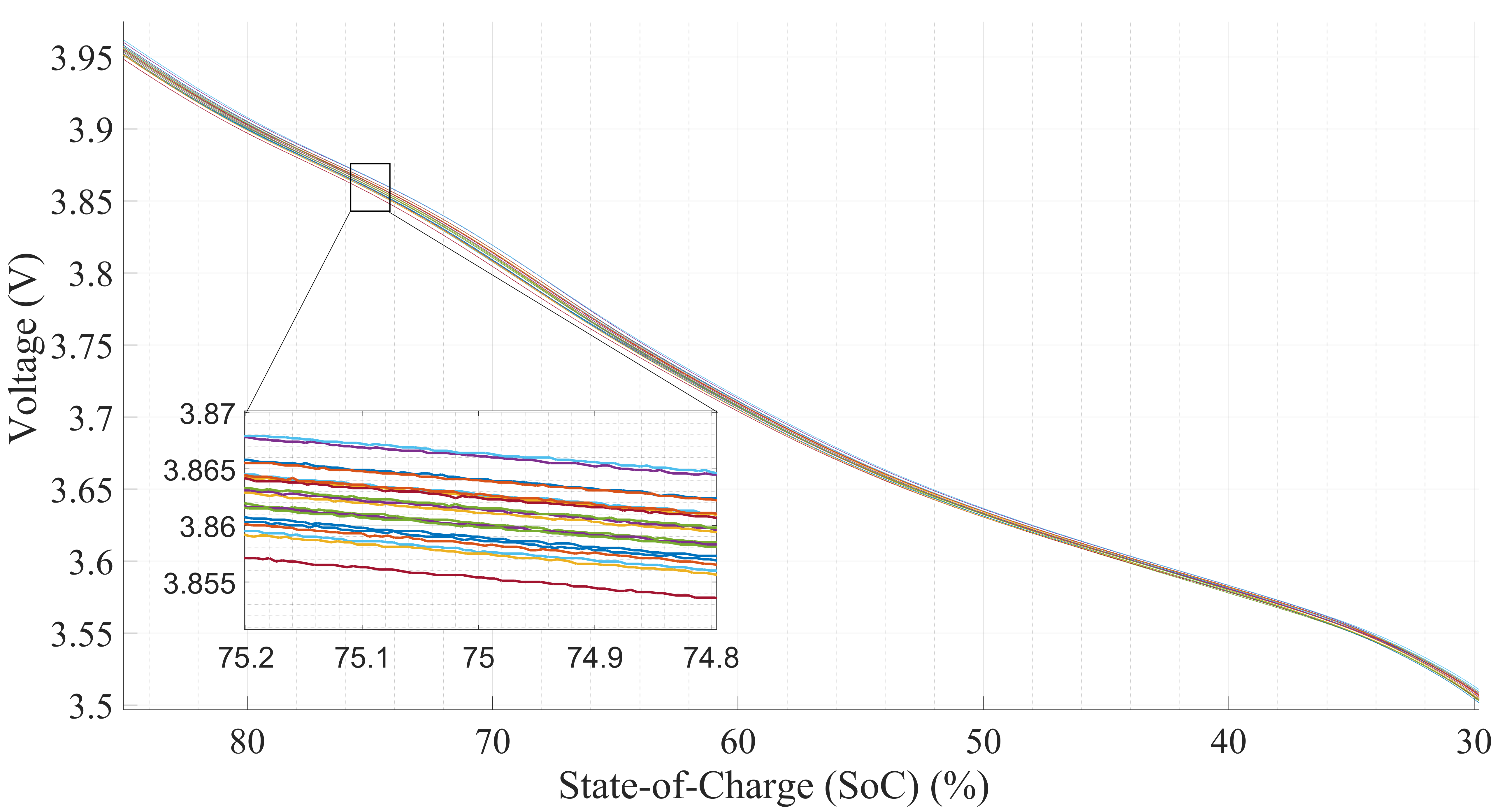}}
\caption{Voltage and SoC variation for $20$ different li-ion cells during their \emph{first} discharge cycle.}
\label{fig:liionSoCvar}
\end{figure}

Our design uses li-ion cells as an emerging energy storage solution in BESS. However, the approach is modular in terms of supporting other battery types at the BESS without modifications to the architecture of DERauth. 
With respect to battery characterization and performance evaluation technologies, electrochemical impedance spectroscopy (EIS) is widely used to monitor changes in batteries under different usage or storage conditions. EIS uses small-amplitude AC signals to measure resistive and capacitive properties over a wide frequency range \cite{EIS}. Despite the popularity of EIS for the evaluation and thus unique identification of battery cells, EIS requires the cell to be disconnected from any load and be at steady equilibrium state. Additionally, measuring an EIS spectrum takes time (often up to many hours) and requires specialized equipment. Hence, it can become prohibitively expensive if applied to multiple cells.

In our design, considering the impracticality of EIS in  real-time applications, we perform BESS profiling using the \emph{model} of li-ion cells which is based on the cell's real-time voltage and SoC. The voltage describes the difference in electrical potential between the poles of a battery. It provides the cell's electromotive force that changes with the SoC according to each particular variant of the li-ion chemistry. The SoC is  the cell's residual charge and therefore the expected operating time. It is a percent ratio of the current battery capacity (measured in $mAh$) and the rated maximum cell capacity \cite{MIT_batt}. 

The behavior profile of li-ion cells regarding voltage and SoC can significantly differ as a result of the intrinsic variability caused by the manufacturing process \cite{edseee.870780620181201}.  Identical li-ion cells manufactured at the same facility and following the same process provide different voltage measurements at the same SoC during their lifetime. The internal resistance, capacitance, and electrochemical effects that cannot be fully controlled during fabrication contribute to these deviations \cite{doi:10.1002/ente.201900201}. Fig. \ref{fig:liionSoCvar}  demonstrates the voltage and SoC discrepancies between $20$ identical li-ion cells during their \emph{first} discharge cycle. As the number of cycles increases, the voltage and SoC variation between different cells of the same BESS, i.e., at the same aging, will further increase. In our design,  the multiple li-ion cells constituting the BESS of the DER are leveraged towards entropy generation, i.e., the voltage and SoC of each battery cell are utilized towards the development of DERauth.

\subsection{System Modules} \label{s:modules}
In this section, we discuss the operating modules of DERauth. The {control} master server and the DER's {embedded} outstation device share a \emph{cell-reply table}. The content of the table is ephemeral and gets updated asynchronously at both ends without requiring secure storage. In the event of compromise, the stored data will become obsolete the moment that the communication round terminates. The {cell-reply} table is defined as $C_{rt}=N\times<c_i, r_i>$. $N$ is the number of BESS cells, and $c_i$ and $r_i$ are the selected BESS cell and its reply, respectively. The length of the cell replies, $r_i$,  depends on the system integrator's constraints and number of 
supported BESS cells. If cell $c_i$ is authenticated, then the reply consists of $r_i$. $C_{rt}$ is initialized at the first authentication request with pseudo-random $r_i$ values and gets modified at both ends after every authentication request.

\begin{figure}[t!]
\centerline{\includegraphics[width=\linewidth]{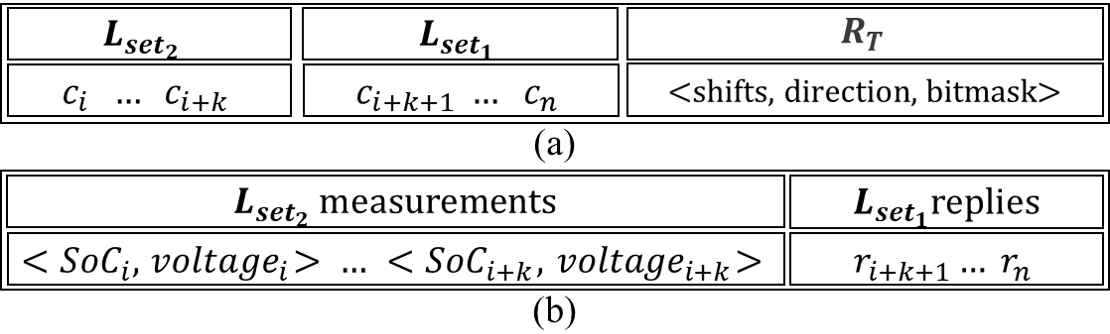}}
\caption{ Layout of (a) challenge and (b) temporary reply $r_{temp}$.}
\label{fig:response}
\end{figure}

At the control server, a \emph{challenge builder module} generates challenges in the form of $<L_{set_{1}}, L_{set_{2}}, R_T>$ as shown in Fig. \ref{fig:response}(a). 
$L_{set_{1}}$ is a set of BESS cells selected to be authenticated and $L_{set_{2}}$ is a set of cells selected to represent the current BESS state, $B_s$, i.e., the real-time SoC and voltage {measurements.} $L_{set_1}$ is not necessarily equal to $L_{set_2}$, and the cells contained in the two sets are arbitrarily defined by the control server with every challenge request. The challenge also includes the reply transformation bits, $R_T$. These bits indicate to the DER's outstation device how the DERauth reply, $r_{auth}$, should be modified before being sent to the master. In our design, $R_T$ values include shift operations and their direction as well as bitmask data. The flexibility of DERauth allows to modify the reply transformation scheme to favor security requirements and effectively increase the replies randomness. We refer to these reversible operations as the \textit{transformation function} $T$ of $R_T$. It allows to generate different CRSeqs for the same pair of $L_{set_1}$ and $L_{set_2}$ which increases the possible challenge-reply space preventing  packet replay or modeling attacks.

\begin{figure}[t!]
\centerline{\includegraphics[width=\linewidth]{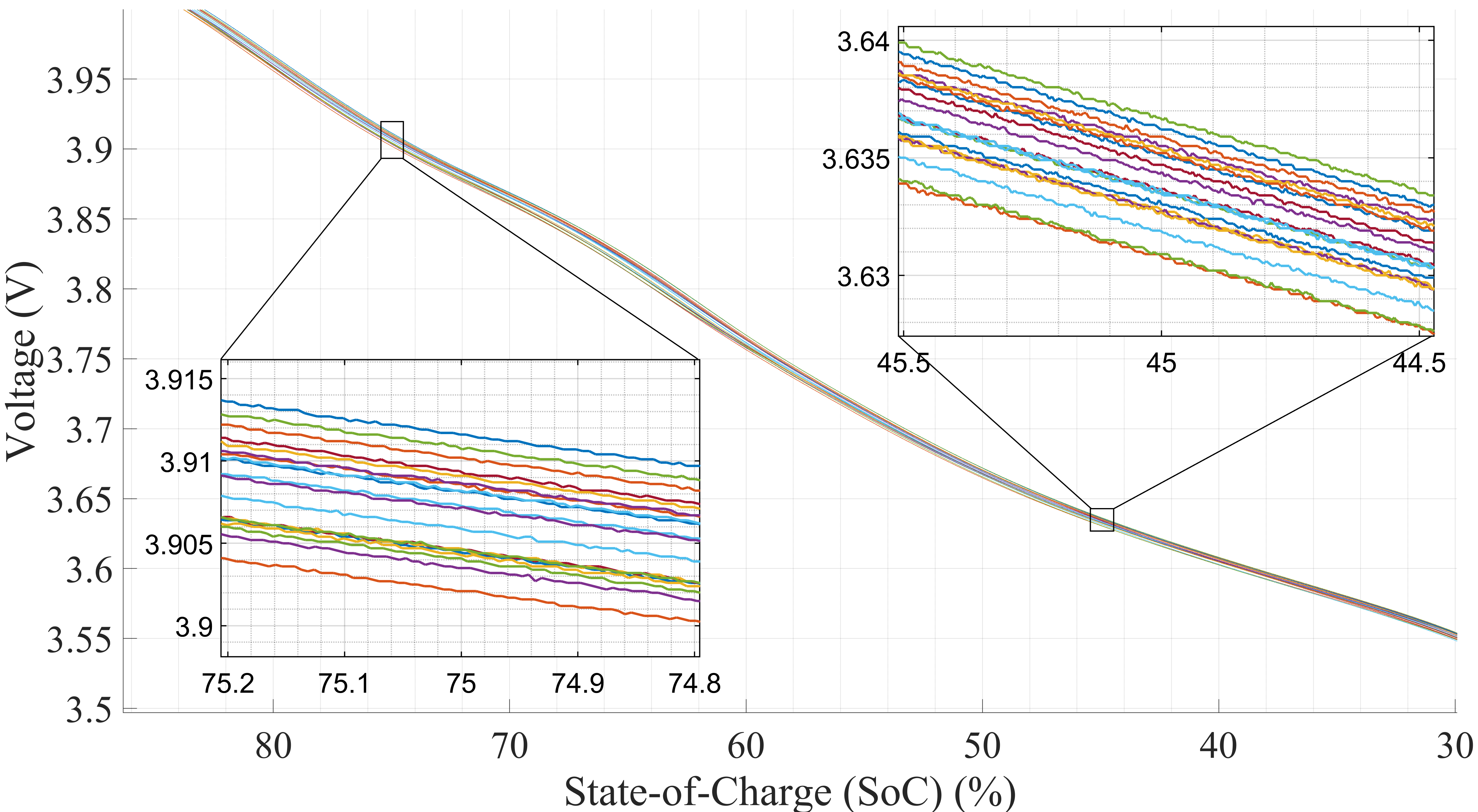}}
\caption{Cycle-to-cycle discharge variations for the same cell.}
\label{fig:VTSSameCell}
\end{figure}

At the DER outstation, there exist three main modules to support the authentication functionality of DERauth. \textit{(i)} A \emph{battery fuel gauge} module which records the real-time SoC and voltage measurements for the requested BESS cells. \textit{(ii)} A \emph{dynamically updating characteristic cell-model (DUCM)}: every BESS cell has its own unique DUCM, a dynamic database with transient SoC measurements and their corresponding voltages for each cell. The relationship between SoC and voltage values is not deterministic as presented in Fig. \ref{fig:VTSSameCell}. The DUCM adjusts cycle-to-cycle and aging variations. The BESS measuring process is constantly running on the outstation unit updating the DUCM with the latest measurements. \textit{(iii)} A \emph{reply builder} module that generates the authentication reply message $r_{auth}$. As part of this module, a temporary reply $r_{temp}$ is constructed and it consists of two parts: (a) the $B_s$ provided by $L_{set_{2}}$ and used to modify the $C_{rt}$ in both the outstation and master, and (b) the replies $r_i$ of cells $c_i$ from $L_{set_{1}}$. Finally, $r_{auth}$ is transformed [$ r_{auth} \equiv T(r_{temp})$] using the transformation function $T$ which can be functionally modified according to application-specific security requirements (Section \ref{s:secDis}).

\begin{algorithm}[t!]
\setstretch{1}
\normalsize
\SetAlgoLined
\DontPrintSemicolon
 exchange $C_{rt}$ \;
 \While{session.valid}{
  challenge.send($L_{set_{1}}, L_{set_{2}}, R_{T} $) \;

  \If{challenge.received}{

      \lIf{compare(measure($L_{set_{1}}$), DUCM)}
      {   \\ \quad replies = get.reply($L_{set_{1}}, C_{rt} $)
          \\ \quad $B_{s}$ = measure($L_{set_{2}}$)
          \\ \quad $r_{auth}$ = transform($B_{s}$, replies, $R_{T}$) 
          \\ \quad update($C_{rt}, B_{s}$), send($r_{auth}$) 
       }
      \lElse{ abort.authentication() }
    }
     \If{reply.received}
     {

       [$B_{s}$, replies] = extract($r_{auth}$) \\
       \lIf{verify(replies, $C_{rt}$)}
       {    \\ \quad update($C_{rt}, B_{s}$), issue.controlCommand()             
       }            
        \lElse{ abort.authentication() }
     }
 }
 \caption{DERauth enrollment and authentication}
 \label{alg:derauth}
\end{algorithm}

\subsection{System Flow}
The design details and the process flow of DERauth are presented in Algorithm \ref{alg:derauth}. The overall design process includes two phases: the \emph{enrollment phase} and the \emph{authentication phase}.

\subsubsection{Enrollment phase}
In this stage, the master server and the DER's embedded outstation device establish a secure communication link to exchange the cell-reply table, $C_{rt}$. The two parties as in a typical power grid environment  communicate from geographically dispersed locations utilizing TCP/IP connections. Protocols such as IEEE 2030.5, Modbus and DNP3 facilitate the majority of these TCP/IP links while the security of the exchanged data is ensured by Transport Layer Security (TLS) (Fig. \ref{fig:protocol_stack}). Based on the established communication link, and following the guidelines of the already-in-place DER protocol, the outstation device shares the $C_{rt}$ with the server.

\subsubsection{Authentication phase}
The master control server sends an authentication challenge, formed by the challenge builder module, to the DER embedded device. Once the challenge is received, the selection of $L_{set_{1}}$ and $L_{set_{2}}$ occurs in accordance to the challenge layout of Fig. \ref{fig:response}(a). 
With the cell selection completed, the outstation device takes SoC and voltage measurements of $L_{set_{1}}$ and $L_{set_{2}}$ using the battery fuel gauge module. In order to authenticate $L_{set_{1}}$, each cell's real-time SoC and voltage are compared to the stored values in its respective DUCM (\emph{self-authentication}). 
Next, DERauth updates $C_{rt}$ based on a predefined operation $p_{[{r_i},{B_s}]}$  
between every reply $r_i$ and BESS state $B_s$ {to avoid storing invariable $C_{rt}$ data at both ends}.  

$r_{auth}$ is then generated by the reply builder module and sent from the DER device to the master server. The server extracts the replies $r_i$ of $L_{set_{1}}$ and the current battery state $B_s$. To achieve that, first the reverse transformation of function $T$ is performed according to the reverse shift and bitmask operations within $R_T$. Then, each cell's $r_i$ is verified using the master server's $C_{rt}$. The $C_{rt}$ at the control server is also updated based on {the} $p_{[{r_i},{B_s}]}$  {operation}. The selection of operations within $p_{[{r_i},{B_s}]}$ and $R_T$ can be adapted based on the time-critical latency requirements of the industrial process and protocol. 

\subsection{Security Discussion}  \label{s:secDis}
In order to address {man-in-the-middle} {(e.g., eavesdropping, packet replay, etc.)} and modeling attacks, we incorporate within the challenge-reply mechanism of DERauth the current BESS state measurements, $B_s$, and the transformation function, $T$. 
The BESS state allows to: \textit{(i)} generate different replies for the same challenges, and \textit{(ii)} update the $C_{rt}$ after every communication round. The inclusion of $B_s$ within DERauth protects against replay and rollback attempts for authentication using old versions of $r_{auth}$. The transformation function, $T$, serves as a data integrity check; any  received reply -- at the master server side -- has to conform to $T$ defined when the challenge was issued. Any modified reply will be disregarded as counterfeit since after reversing $T$ the data will have an incorrect format. 

According to the design of DERauth, all of the three following conditions are necessary for an attacker to compromise DERauth's operation: \textit{(i)} exact knowledge of the challenge and reply layout as well as their content representation, \textit{(ii)} the transformation function $T$ being used, and \textit{(iii)} the most current version of the $C_{rt}$. The first requirement is necessary in order to be informed which cells are measured for the $B_{s}$ and which are authenticated against their DUCM. The second condition is required in order to properly structure the reply and not get rejected (as modified) from the master server. Overall, the transformation function can be changed every time an enrollment phase is initiated; its update frequency can be specified by the DERauth integrator based on the security and flexibility constraints of the monitored DER asset. The third requirement is needed in order to provide the correct reply values $r_i$ for the indicated $L_{set_{1}}$ cells, else the authentication will fail. Knowledge of the current version of $C_{rt}$ requires physical access to the master server or the outstation device. Also,  since $C_{rt}$ gets updated after every communication round, any previous knowledge becomes ineffective. 

\section{Experimental Results} \label{s:Implementation And Evaluation}

\subsection{Design Parameters} 
For the implementation of DERauth we opted for $64$-bit CRSeqs. The $c_i$ and $r_i$ of $C_{rt}$ are selected to be $8$-bit long, and $N$ was set to $100$ by extrapolating the measurements of multiple BESS cells. 
Based on the size of $c_i$, we use in the challenges: four cells for $L_{set_1}$ and two cells for $L_{set_2}$ (Fig. \ref{fig:response}(a)). The remaining $16$ bits in the challenge layout are reserved for the reply transformation, $R_{T}$.  The first five bits of $R_{T}$ define the number of positions that each reply will be shifted (cyclically), the following three indicate the direction of the shift, and the remaining eight bits are an XOR bitmask applied to the shifted reply. {We adopt bitwise logical operations and shifts for the transformation function $T$ as well as the $p_{[{r_i}{B_s}]}$ operation used for the $C_{rt}$ updating process, ensuring that DERauth meets the stringent timing requirements of industrial protocols such as DNP3-SA. }
Regarding the replies $r_{temp}$, their first half is the $B_s$ of $L_{set_{2}}$ and the second half includes the $8$-bit $r_i$ of $L_{set_1}$, as shown in Fig. \ref{fig:response}(b).

\subsection{Implementation Setup} \label{s:experiment_specifics}
\begin{figure}[t!]
\centering
    \subfloat[]{
            \includegraphics[width=0.65\linewidth]{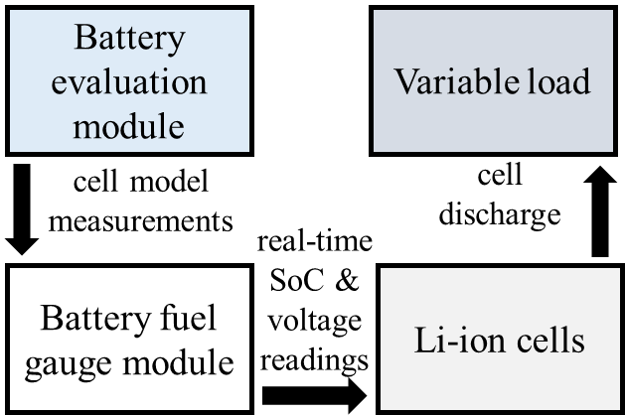} 
            \label{subfig:schematic}
    }\\ 
   \subfloat[]{
            \includegraphics[width=0.65\linewidth]{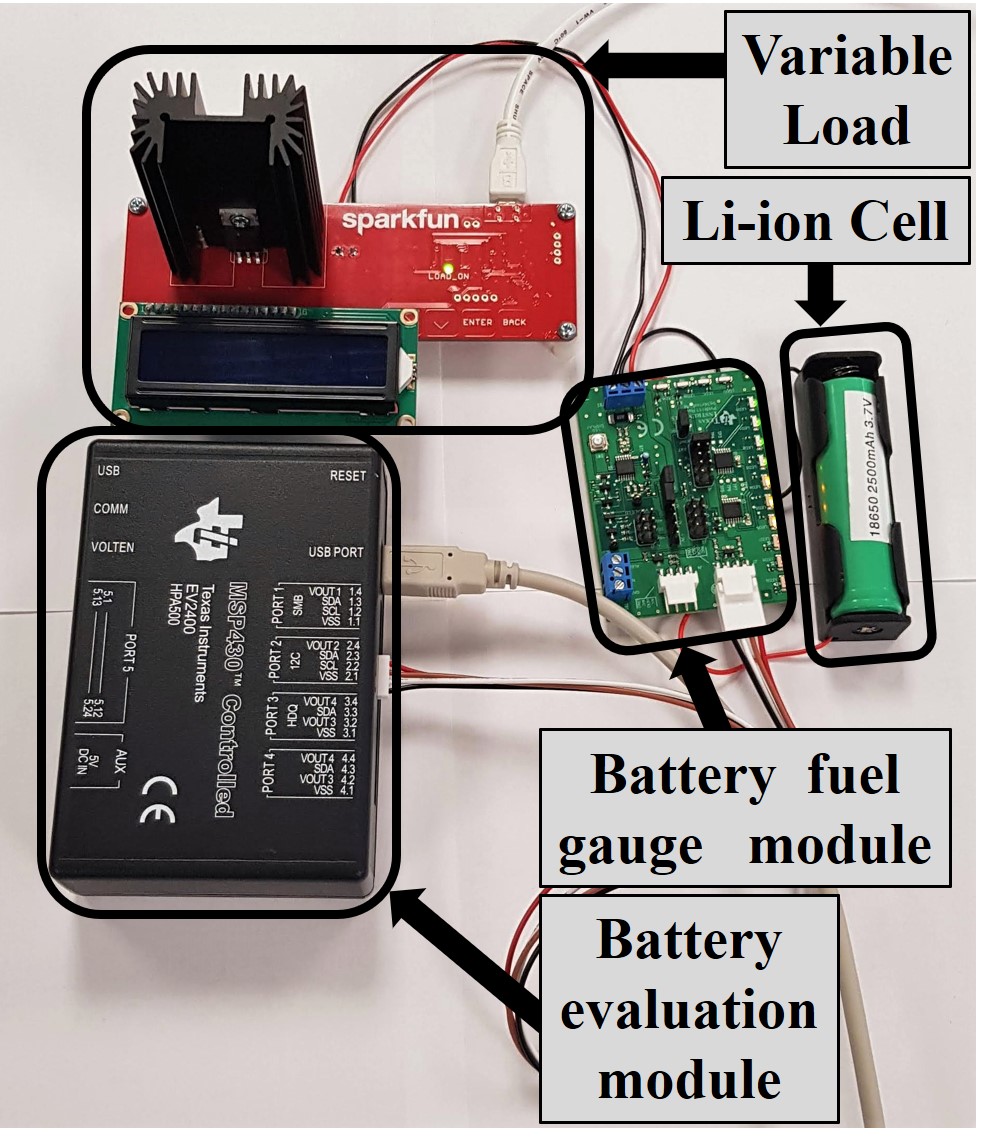}
            \label{subfig:eval_prototype_image}
    }
\caption[f]{\subref{subfig:schematic} Schematic diagram and \subref{subfig:eval_prototype_image} experimental measurement setup.} 
\label{fig:eval_prototype}
\end{figure}

In order to evaluate DERauth, we have developed a proof-of-concept implementation in which li-ion battery cells (Samsung 25R 18650) are used as the BESS of DER. For the cell discharge characterization, which provides the cell SoC and voltage, we assume a constant load of $500$ $mA$. This is sufficient since DUCM constantly monitors the BESS cells and can account for varying load conditions. For the extraction of DUCM measurements, we use the battery evaluation module TI EV2400 with the battery fuel gauge TI BQ34Z100EVM. The setup provides the SoC and voltage measurements of cells with a maximum error of $\pm{1}\%$. To attain the highest possible accuracy from the {battery} fuel gauge, each {battery} cell is subjected to a ``learning cycle''. The cycle involves a full discharge for the fuel gauge module to accurately calculate each cell's impedance. The setup is presented in Fig. \ref{fig:eval_prototype}.

As presented in Section \ref{s:methodology}, during the authentication phase the outstation device compares the measured SoC and voltage for $L_{set_1}$ cells with their corresponding DUCM. The DUCM consists of the discharge measurements between voltage levels of $\Delta V =$ $4$  -- $3.45$ $V$ 
as the nominal operating range per li-ion cell \cite{leng2015effect}. Exceeding those limits could severely impact the cell's condition and performance (e.g., low capacity, poor cycle life, high self-discharge, etc.). {For example, a prolonged low voltage may cause dissolution of metals (e.g., copper) while a high voltage can cause battery degradation.} SoC and voltage {measurements} are acquired at $10$ $mAh$ intervals for every cell. Obtaining and storing these values more frequently (e.g., $1$ $mAh$) has also been examined. However, this offers marginal  accuracy improvements while the memory footprint increases almost tenfold. In the scenario of additional memory constraints, the measurement voltage interval $\Delta V$ could be further decreased, assuming the BESS is operating in the $\Delta V$ range, without altering the  {operating} principles of DERauth or impacting the authentication reliability.

\begin{figure}[t!]
\centerline{\includegraphics[width=\linewidth]{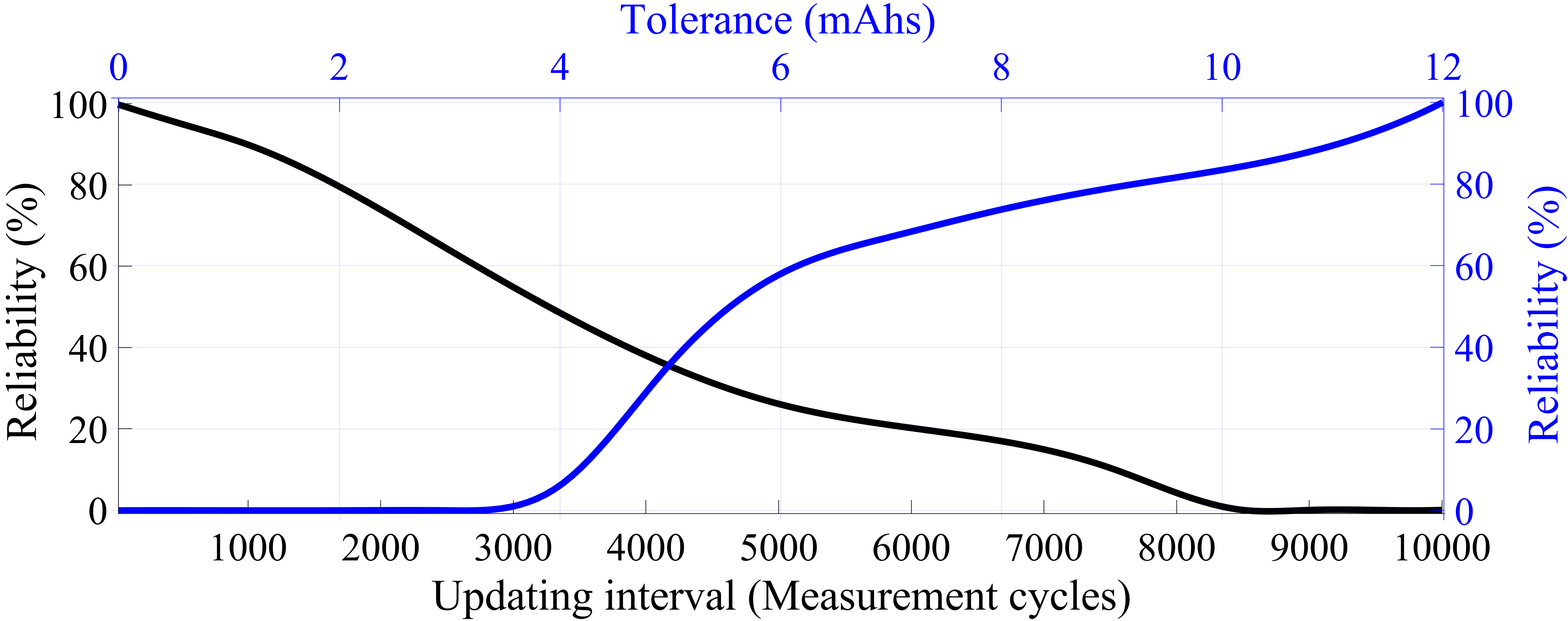}}
\caption{Reliability Vs. updating interval (black) and tolerance threshold (blue).}
\label{fig:rel_tol_up}
\end{figure}

\subsection{Evaluation Analysis}
In this part, we {present the evaluation of our implementation}. We 
describe the procedure to ensure that cell SoC and voltage measurements are sufficient for authenticating each li-ion battery cell. We also discuss DUCM accuracy and present the experimental results that illustrate the impact of the DUCM updating frequency. In addition, we assess the resiliency of the DERauth's CRSeqs against machine learning attacks. We test the CRSeqs with three different multilayer perceptron neural networks (MLP-NN).

\subsubsection{Battery model evaluation}

In order to examine the battery model and the reliability of DERauth, we measure ${S_{auth}}/{A_{auth}} \times 100\%$, where $S_{auth}$ is the number of successful authentications and $A_{auth}$ is the total {number of} attempted authentications for the same  cell. In order to construct the battery model, we discharged cells for $20$ cycles. Fig. \ref{fig:VTSSameCell} shows that, even for the same cell, discharge curves vary from cycle-to-cycle depending on inherent battery features (e.g., electrochemical properties, resistance, capacitance, process variations, etc.), or exogenous causes such as the  the cell's operating temperature, the type of load connected to the cell, aging effects, state of health, etc. A \emph{static} battery model (e.g., Randle's circuit) could not address the time varying characteristics of the cells \cite{randle}. DUCM is constantly polling the SoC and voltage of every cell and contributes to an updated cell model.

Towards addressing the cycle-to-cycle discharge variation, we define a tolerance threshold ($\tau$) representing the difference between real-time state of $L_{set_1}$ and DUCM. The blue curve of Fig. \ref{fig:rel_tol_up} shows how the selection of $\tau$ affects the reliability of the cell authentication \emph{before the deployment of DUCM}. A higher threshold could result in efficiently self-authenticating the cells under test due to the smaller disparity between the cells' real-time measurements and their corresponding model when compared against $\tau$. As the threshold increases, although we can reliably self-authenticate any requested cell, we can also have many false positives, i.e., cells at similar SoC and voltage values (\emph{cell model}), which could get authenticated if the discrepancy between their model and the real-time measurements is less than $\tau$. This could allow adversaries to acquire cell replies, $r_{i}$. DUCM minimizes $\tau$ to $1$ $mAh$ while authenticating the defined cells with reliability of $90\%$, eliminating the false-positives issue. DUCM achieves that by updating the model of each  cell -- to account for the offset between previous cycles and aging over the cell's lifetime -- keeping it always below the tight threshold $\tau$.

\begin{figure}[t!]
\centering
    \subfloat[]{
            \includegraphics[width=0.85\linewidth]{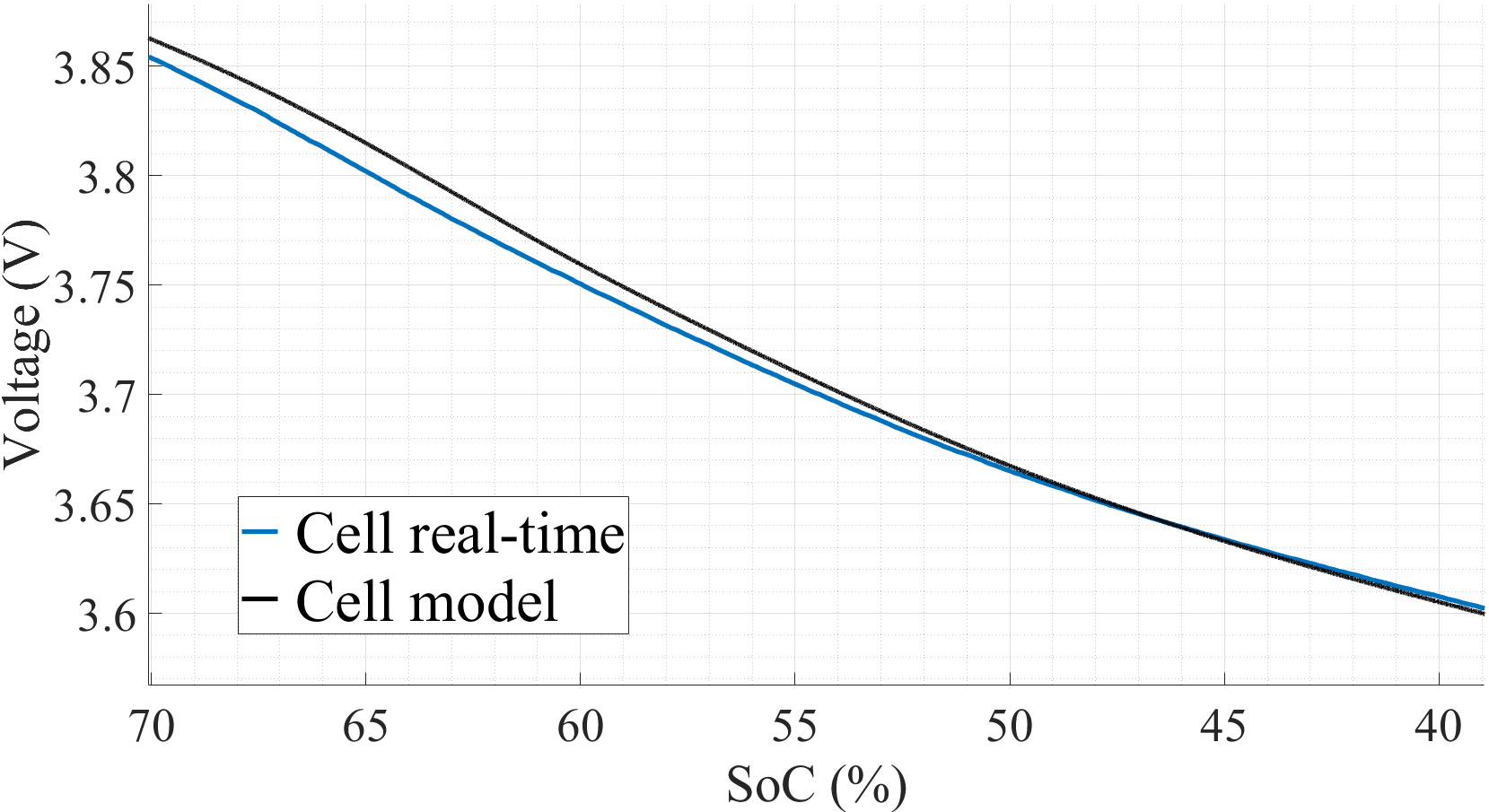}
            \label{fig:subfig1_dyn_update}
    }\\
   \subfloat[]{
            \includegraphics[width=0.85\linewidth]{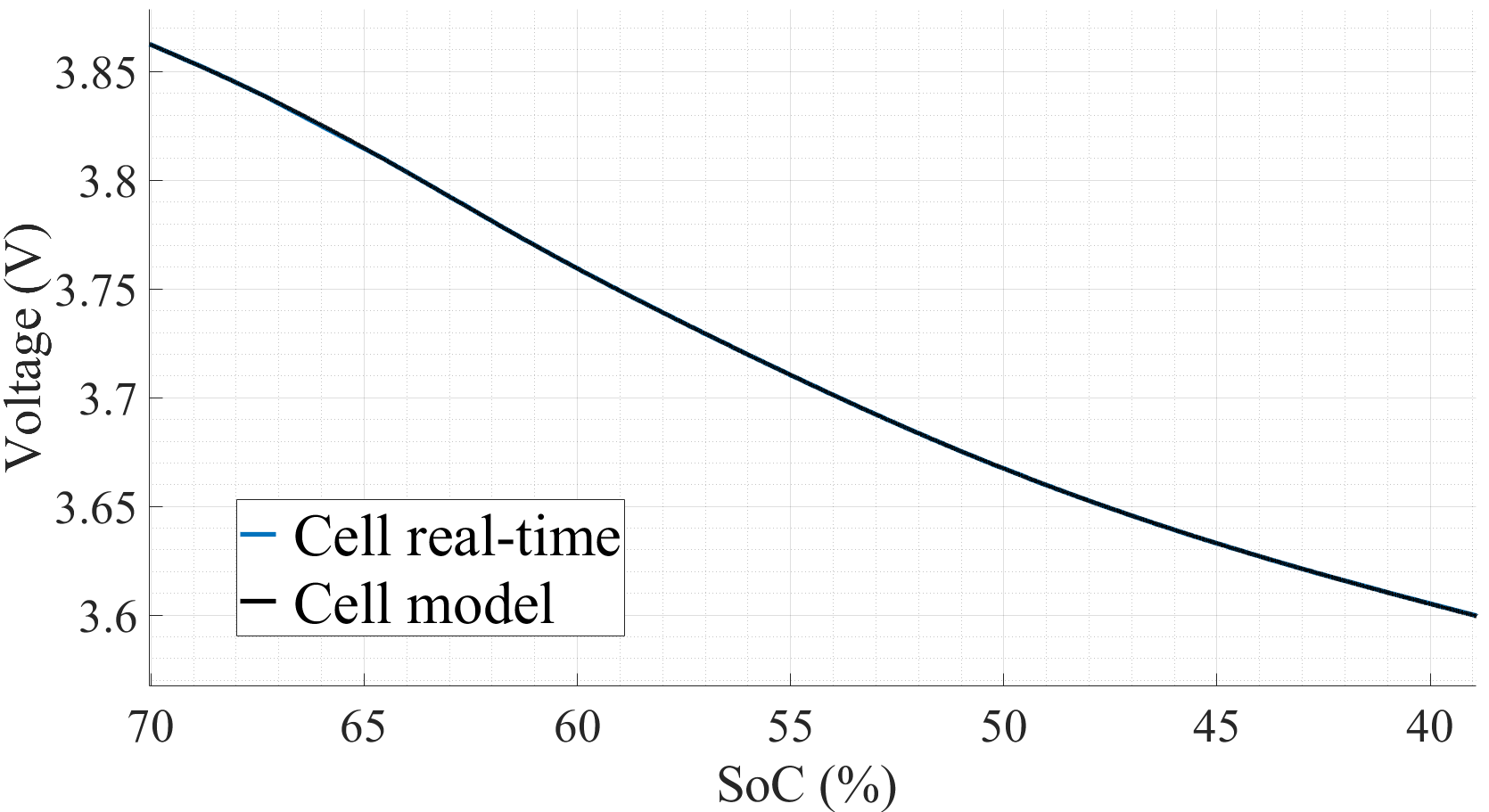}
            \label{fig:subfig2_dyn_update}
    }

\caption[CR]{Cell real-time Vs. cell model discharge behavior. \subref{fig:subfig1_dyn_update} without the dynamically updating characteristic cell-model (DUCM), and \subref{fig:subfig2_dyn_update} with 
DUCM.}
\label{fig:dyn_update}
\end{figure}

Fig. \ref{fig:VTSSameCell} presents the difference in discharge graphs between $20$ discharge cycles. Without DUCM, the cell model would be identical to the first discharge cycle (cycle $1$), thus real-time and cell model values would significantly differ over time. This is presented in Fig. \ref{fig:dyn_update}(a). DUCM can eliminate the cycle-to-cycle discrepancy as illustrated in Fig. \ref{fig:dyn_update}(b). The DUCM accurately tracks the transient discharge behavior of every cell {and addresses the cycle-to-cycle variations}. We further examine DUCM reliability with $10k$ measurements taken every $2$ $secs$ (measurement cycle). Fig. \ref{fig:rel_tol_up} illustrates how the updating interval of DUCM affects authentication reliability. For example, if the cell model is updated every $1,000$ measurements ($10$ times in $10k$ measurements) the authentication reliability is $89.92$\%.

\subsubsection{System Evaluation}
Physical system variations can be utilized for lightweight authentication protocols used by {resource-constrained platforms}. However, the challenge-reply mapping could be predicted if a number of CRSeqs is available. For instance, prediction accuracy that exceeds 90\% with a small number (in the range of a few thousands) of CRSeqs has been reported  \cite{delvaux2019machine}. In order to evaluate DERauth against such modeling attacks resulting from machine learning, we provide a set of CRSeqs to a MLP configuration \cite{MLP}. Increasing the number of MLP layers creates well-defined relationships between inputs and outputs improving the learning efficacy of the algorithm. In our case, three different network architectures are utilized. Their best prediction results are achieved with a dataset of 100$k$ CRSeqs. The 100$k$ CRSeqs are generated in Python as part of the challenge-reply protocol implementation. As for the entropy of CRSeqs, we utilize real-time battery voltages and SoC values collected by our experimental setup. Different sizes of hidden layers are used in order to enhance the learning rate of the {machine learning modeling} attacks. A grid search is also performed to fine-tune the hyperparameters of the MLPs.  In Table \ref{tab1}, we provide results for the three MLP-NN architectures with different batch sizes for the training stage of the MLP-NN. 

\begin{table}[t]

    \caption{MLP Neural Network Performance  with 100k CRSeqs.}
\small	
    \begin{center}
    \begin{tabular}{||c|c|c|c||}
    \hline \hline
    \textbf{Network} & \textbf{Batch}& \textbf{Validation}& \textbf{Prediction}\\ 
    \textbf{Architecture} & \textbf{size}& \textbf{accuracy}& \textbf{accuracy} \\
    \hline \hline
    $64-320-128-64-64$ & 32 & 81.50\% & 81.46\%  \\
    \hline
    
    \hline
    $64-320-128-64$ & 32 & 81.90\%  & 80.93\%  \\
    \hline
    
    \hline
    $64-64-64-64$ & 64 & 71.47\% & 71.26\%  \\
    \hline \hline
    \end{tabular}
    \label{tab1}
    \end{center}
\end{table}
\normalsize

\begin{figure}[t]
\centerline{\includegraphics[width=\linewidth]{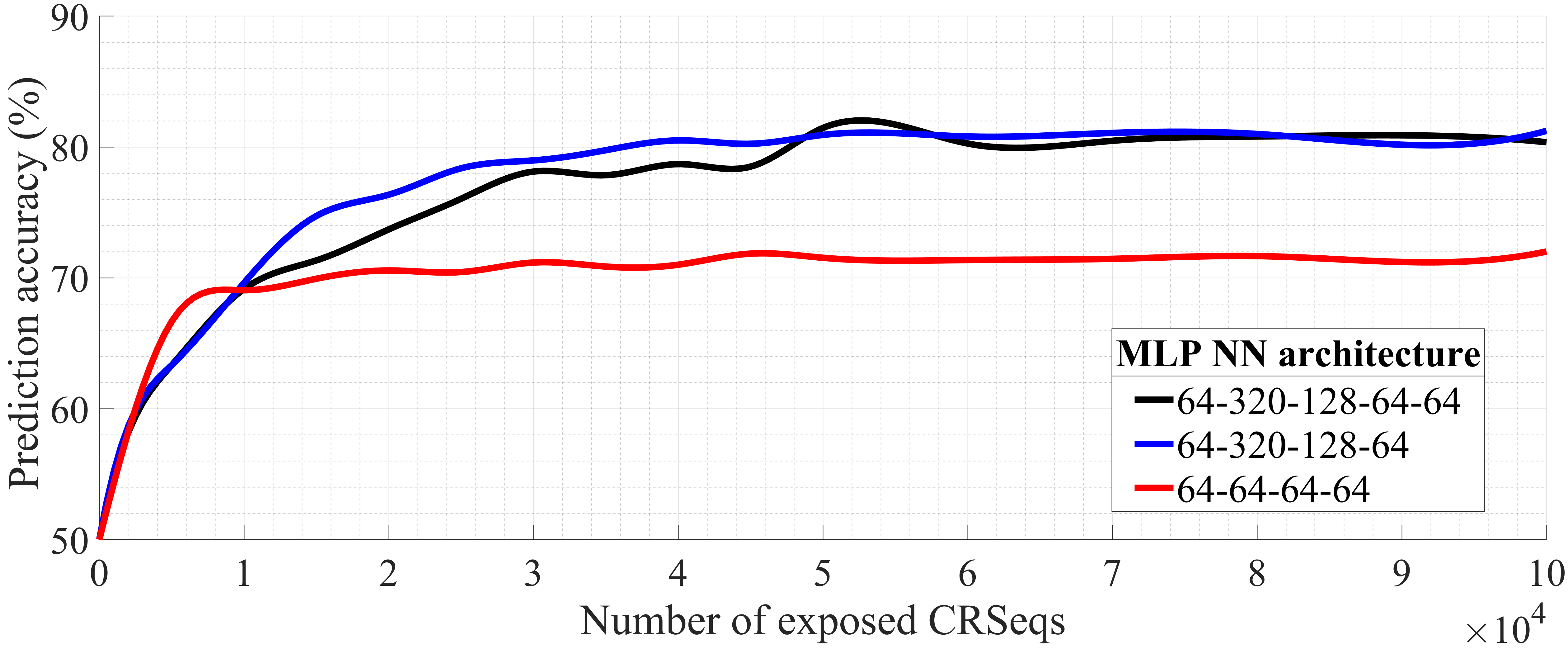}}
\caption{Prediction accuracy Vs. known CRSeqs datasets.}
\label{fig:predictionACC}
\end{figure}

We train the configurations of the MLP-NN algorithms using different dataset sizes ranging from 1$k$ to 100$k$ CRSeqs and a 20\% validation split. The 64-bit CRSeqs emulate the behavior of DERauth in a real-world scenario. The machine learning attack model is tested on generated sets using the same procedure. Our results indicate that increased number of available CRSeqs ($>$ 10$k$) has minor improvements in prediction accuracy. This emphasizes the robustness of our proof-of-concept. Ideally, the accuracy should be 50\% (complete randomness). In our experiments, the MLP never exceeds accuracy levels of $\approx$ 82\% despite increasing the size of the CRSeq training set. This level could be further decreased using longer CRSeqs (e.g., 128-bit) able to incorporate higher BESS entropy (more cell replies $r_{i}$ and $B_{s}$ measurements). The results showing the prediction accuracy for the three different configurations of MLP networks are depicted in Fig. \ref{fig:predictionACC}.

\section{Conclusion and Future work}\label{s:Conclusion}
This paper presents DERauth, a proof-of-concept hardware authentication framework for DERs in power grids using li-ion battery cells. This prototype version shows the potential of leveraging existing hardware to authenticate BESS-based DER systems. Thus, DERauth can serve as an add-on feature in existing industrial DER protocols. We have developed a dynamic process to update the BESS state and evaluate our model using MLP-NN. Building an ecosystem for a scalable and modular design will be included in our future work. Our research will incorporate a formal security analysis, fuzzy extractor error correction techniques, hardware-in-the-loop experiments, 
investigation of longer CRSeqs to explore trade-offs between security, performance and storage overhead, and variable loads to simulate cells discharge behavior. Furthermore, we will develop the required application program interface (API) which will enable the integration of our authentication framework into existing communication protocol standards (e.g., DNP3, IEEE 2030.5, etc.) used for issuing commands between aggregators and DER devices.

\section*{Acknowledgement}
Partial support of this research was provided by the Woodrow W. Everett, Jr. SCEEE Development Fund in cooperation with the Southeastern Association of Electrical Engineering Department Heads.

\bibliographystyle{IEEEtran}
\bibliography{ISVLSI_main}

\end{document}